\documentstyle[12pt]{article}
\newcommand{\be}{\begin{equation}}
\newcommand{\ee}{\end{equation}}
\newcommand{\bear}{\begin{eqnarray}}
\newcommand{\ear}{\end{eqnarray}}

\newcommand{\slp}{\raise.15ex\hbox{$/$}\kern-.57em\hbox{$p$}}
\newcommand{\slv}{\raise.15ex\hbox{$/$}\kern-.57em\hbox{$v$}}
\newcommand{\slet}{\raise.15ex\hbox{$/$}\kern-.57em\hbox{$\eta$}}
\newcommand{\slB}{\raise.15ex\hbox{$/$}\kern-.57em\hbox{$B$}}
\newcommand{\slb}{\raise.15ex\hbox{$/$}\kern-.57em\hbox{$b$}}
\newcommand{\slW}{\raise.15ex\hbox{$/$}\kern-.57em\hbox{$W$}}
\newcommand{\GeV}{\rm GeV}
\newcommand{\eV}{\rm eV}

\begin{document}
\begin{titlepage}
\begin{flushright}
HD-THEP-99-15
\end{flushright}
\quad\\
\vspace{1.8cm}

\begin{center}
{\Large A simple scheme for masses and mixings of quarks and neutrinos}\\
\vspace{2cm}
Berthold Stech\footnote{e-mail: B.Stech@thphys.uni-heidelberg.de}\\
\bigskip
Institut  f\"ur Theoretische Physik\\
Universit\"at Heidelberg\\
Philosophenweg 16, D-69120 Heidelberg\\
\vspace{3cm}

\end{center}

\begin{abstract}
The mass matrices of charged fermions have a simple structure if expressed
in powers of the small parameter $\sigma=(m_c/m_t)^{1/2}$.
It is suggested that the mass matrix of the three heavy neutrinos
occuring in grand unified theories can be expressed in terms of the same
parameter. The requirement that these heavy neutrinos carry
different $ U(1) $ generation quantum numbers gives rise to an almost
unique form for this matrix. By applying the see-saw mechanism, the
mass splitting of the two lightest neutrinos comes out to be tiny,
favoring the vacuum oscillation solution for solar neutrinos.
The mixing matrix is of the bimaximal type but contains also
CP violating phases.

\end{abstract}
\end{titlepage}

\newpage
\section{Introduction}
It is well known that the masses and mixings of quarks show a hierarchical
pattern. The mass matrices, the objects of the theoretical description,
can be expressed in terms of powers of a small parameter with coefficients
of order one \cite{1}. Since among the quarks the top quark plays
an outstanding role, a natural choice for the small parameter is the
quantity $\sigma=(m_c/m_t)^{1/2}$ \cite{2}. With this
choice the mass ratios $m_u:m_c:m_t$ taken at a common scale
are simply $\sigma^4:\sigma^2:1$. Recent indications for solar \cite{3}
and atmospheric \cite {4} neutrino oscillations, and thus for non-zero
neutrino masses, raise the question of the structure of the mass matrices
for leptons. Are they related to the mass matrices of quarks?

In this article I argue for an intimate connection between quark and
lepton masses and mixings. The general suggestions from grand
unified theories are used: the see-saw mechanism  and  the expected
near equality of the Dirac-neutrino mass matrix with the up-quark
mass matrix. For the mass matrix of the heavy neutrinos a new
hypothesis is needed. I assume that each heavy neutrino carries
a quantum number of a new $U(1)$ symmetry which governs
the leading powers of the small parameter $\sigma$ occurring
in this mass matrix. With the single condition that these ``charges''
are not zero one gets strong restrictions for the form of this mass matrix.
The consequences for the light neutrinos are drastic: i) the mass
splitting of the two lightest neutrinos is tiny and favors
the vacuum oscillation solution for solar neutrinos, and ii) the
mixing matrix is close to the one for bimaximal mixing.

\section{Quarks}

Today the masses and mixings of quarks are reasonably well known with
only the exception of the CP-violating phase parameter.
After the choice of a convenient basis the mass matrices for up and
down quarks can be written down. For instance, one can use as in
\cite{2} a real and symmetric matrix $m_U$ for the up quarks and a
hermitian matrix $m_D$ for the down quarks such that the elements 
$(m_U)_{11}$, $(m_D)_{13},(m_D)_{31}$ and $(m_D)_{23},(m_D)_{32}$
are strictly zero. This basis has the advantage that the only complex matrix
element is $(m_D)_{12}=(m_D)^*_{21}$. More important, by expressing the
matrix elements in powers of the small quantity $\sigma=(m_c/m_t)^{1/2}
\simeq0.058\pm0.004$ each independent matrix
element  has a different power of $\sigma$ for the up as well as the 
down quark matrices
with factors of order 1. Our present 
knowledge on masses and mixings is compatible with the
results obtained from the mass matrices \footnote{Compared to ref. \cite{2},
the mass of the strange quark is taken to be $m_s \approx
\sigma/3\ m_b$ instead of $\approx \sigma/2\ m_b$ and also $m_d/m_b \approx 6 \sigma^3$. The smaller values seem
to be more appropriate.} (taken at the mass scale
of the vector boson $Z$):

\be\label{1}
m_U=\left(\begin{array}{ccc}
0&\sigma^3/\sqrt2&\sigma^2\\
\sigma^3/\sqrt2&-\sigma^2/2&\sigma/\sqrt2\\
\sigma^2&\sigma/\sqrt2&1\end{array}\right)m_t,\quad m_D=
\left(\begin{array}{ccc}
0&-i\sqrt2\sigma^2&0\\
i\sqrt2\sigma^2&-\sigma/3&0\\
0&0&1\end{array}\right)m_b\quad\ee
and $\sigma=0.057$.
 For the purpose of this paper in
which we want to use the up quark mass matrix also for neutrinos,
it is convenient, however,
to transform to a basis in which $m_U$ is diagonal. To leading order
in $\sigma$ one gets 
\be\label{2}
m_U=\left(\begin{array}{ccc}
\sigma^4&0&0\\
0&\sigma^2&0\\
0&0&1\end{array}\right)\quad m_D=\left(\begin{array}{ccc}
O(\sigma^3)&-i\sqrt2\sigma^2&i \sigma^2\\
i\sqrt2\sigma^2&-\sigma/3&i \sigma/\sqrt2\\
-i\sigma^2&-i\sigma/\sqrt2&1\end{array}\right)m_b\quad.\ee

Clearly, the simple factors in front of the powers of
$\sigma$ in (\ref{1}) are guessses and have to be changed, 
or higher order terms 
in $\sigma$ have to be included, when more precise information
on masses and mixings become available. Also, for definitness, 
``maximal CP-violation'' has been assumed.
It is defined to maximize the area of the unitarity triangle
with regard to changes of the phases of the off-diagonal elements
keeping their magnitudes fixed. Maximum CP-violation defined
this way allowed us to bring the off-diagonal elements of $m_D$
into the form of an antisymmetric hermitian matrix \cite{5}.
Within the accuracy of only a few degrees one obtains a right-handed
unitarity triangle with angles
$\alpha\approx 70^o,\quad \beta\approx 20^o,\quad \gamma
\approx 90^o.$
Independent of the phases 
the mass matrices (\ref{1}) demonstrate that masses and mixings
are governed by the same small parameter in a simple fashion.
With $\sigma=0.057$ the numerical values for the quark mass ratios
at the common scale $m_Z$
and the absolute value of the Cabibbo-Kobayashi-Maskawa matrix
CKM as obtained from (\ref{1}) are
\be\label{3}
\frac{m_u}{m_t}=1.1\cdot10^{-5},\quad
\frac{m_c}{m_t}=3.2\cdot10^{-3},\quad
\frac{m_d}{m_b}=1.1\cdot 10^{-3},\quad
\frac{m_s}{m_b}=2.0\cdot10^{-2},\ee
\be\label{4}
Abs[CKM]=\left(\begin{array}{lll}
0.97&0.22&0.003\\
0.22&0.97&0.040\\
0.010&0.039&1\end{array}\right)~.\ee

\section{Neutrinos and Charged Leptons}

The recent indications for neutrino oscillations imply
finite neutrino masses and lepton number violation. 
For a thorough discussion on possible scenarios and for 
the relevant literature I refer to ref. \cite{6}. Some 
approaches based on grand unified theories can be found 
in \cite{7,8}. Here, I will take suggestions
from grand unified theories without specifications of 
the group and Higgs representations: The standard model 
is extended by adding
three two-component neutrino fields $\hat\nu_e, \hat\nu_\mu,
\hat\nu_\tau$ which are singlets with respect to the standard
model gauge group. Since the masses of these fields are not
protected, the total $6\times 6$ neutrino mass matrix has a block
structure consisting of a $3\times3$ matrix $M$ with very large
entries and a Dirac-type  mass matrix $m_\nu^{Dirac}$ which connects
the light with the heavy fields. At the scale of the heavy neutrinos
one expects a close
connection between $m_{Dirac}$ and the charged lepton mass
matrix $m_E$ with the up-quark mass matrix and the down-quark mass
matrix, respectively \cite{5}. For a non-singular matrix $M$ the light
neutrinos become Majorana particles according to the see-saw
mechanism. Their mass matrix $m_\nu$ is given by
\be\label{5}
m_\nu=-m_\nu^{Dirac}\cdot M^{-1}\cdot (m_\nu^{Dirac})^T~~.\ee
In the following I use the relation
\be\label{6}
m^{Dirac}_\nu=m_U\ee
and postpone a remark on possible deviations to the end of section 5.

Because the top quark mass is so large compared to all other quark
masses, it is convenient to take a basis in which $m_U$ is diagonal
as already done in (\ref{2}). A particular interesting connection 
between quarks and neutrinos will exist if besides $m_U$ and 
$m^{Dirac}_{\nu}$ 
also the mass matrix $M$ has a simple structure in this basis and 
the parameter $\sigma$ plays there a similar role. 
I will explore this possibility.

Let us therefore express the entries of $M$ in terms of powers of
$\sigma^2$. To give significance to such a form it should be possible
to assign $U(1)$ generation charges to the heavy neutrino fields. 
Generation charges can be decisive for determining the structure 
of mass matrices, see e.g. \cite{7,9}. To restrict these charges 
I will require that the three $U(1)$ quantum numbers differ from 
each other, are not zero and that not all elements of $M$ vanish 
in the limit $ \sigma \rightarrow 0 $. 
As a consequence, two of the three fields must carry opposite charges, 
and $M$ provides for $\sigma \rightarrow 0$ a mass term of the Dirac type for 
a heavy neutrino, i.e. a neutrino described by two different 
two-component fields.
The mass matrix $M$ which satisfies the requirement and has the
entries surviving for $\sigma\to 0$ at the  most symmetric place, 
i.e. $\hat\nu_e$ has the opposite charge of $\hat\nu_\tau$,
has the structure
\be\label{7}
M=\left(\begin{array}{ccc}
\sim\sigma^6&\sim\sigma^2&1\\
\sim\sigma^2&\sim\sigma^2&\sim\sigma^4\\
1&\sim \sigma^4&\sim\sigma^6\end{array}\right)M_0~~.\ee
The $U(1)$ charges of $\hat\nu_e,\hat\nu_\mu,
\hat\nu_\tau$ are $-3/2~,~1/2~,~3/2~$,
respectively; they determine the powers of $\sigma^2$. 
If we dismiss matrices which have determinant zero when 
neglecting higher orders than $\sigma^2$, 
the form (\ref{7}) is unique
apart from a reflection on the cross diagonal corresponding to
the charges $-3/2~,~-1/2~,~3/2~$.
As in the case of
the matrix $m_U$, the unknown factors in (\ref{7})
should be of order 1~. In particular, if there is a close correlation 
with $m_U$, the factor of $\sigma^2$ in the first row and first 
column ($p$ in eq. (\ref{8})) should be equal or very close to one. 
Because of the smallness of $\sigma^4, \sigma^6$
$M$ can be used in the simpler form
\be\label{8}
M=\left(\begin{array}{ccc}
0&p\sigma^2&1\\
p\sigma^2&r\sigma^2&0\\
1&0&0\end{array}\right)M_0~~.\ee
One can check that the approximation (\ref{8}) is also applicable when 
calculating $m_\nu$
according to (\ref{5}), (\ref{6}) even though the inverse of the matrix
$M$ enters there. Moreover, a simple consideration of the original
$6\times 6$ neutrino mass matrix (with zero entries in the light-light
sector) shows that the coefficients $p$ and $r$ can be taken to
be real.

For the mass matrix of the light neutrinos, eqs. (\ref{5}-\ref{8}) give
\be\label{9}
m_\nu=-\left(\begin{array}{ccc}
0&0&r\sigma^2\\
0&1&-p\\
r\sigma^2&-p&p^2\end{array}\right)\frac{\sigma^2}{r}
\frac{m_t(M_0)^2}{M_0}~~.\ee
The neutrino mass spectrum obtained from this mass matrix
is interesting in view of the recent neutrino data. Taking
$r=p=1$ and adjusting $M_0$ such that the largest eigenvalues
$(m_3)$ becomes $m_3 \approx 0.055$ eV, one gets
$M_0 \approx 10^{12}\ {\GeV}, \ m^2_3-m^2_1 \approx 3\cdot
10^{-3}\ ({\eV})^2$ and $m^2_2-m^2_1 \approx 10^{-11}({\eV})
^2$ . Furthermore, the neutrino mixing matrix
obtained from (\ref{9}) with $p=1$ shows the bimaximal mixing 
discussed in \cite{10}. Thus, the neutrino mass matrix (\ref{9}) 
favors large mixing angles for the atmospheric and for the solar 
neutrinos, and the vacuum oscillation solution \cite{11} for 
the latter. But the neutrino mass matrix obtained here seems 
not compatible with the indications for 
$\hat\nu_{\mu} \rightarrow \hat\nu_e $ oscillation reported by 
the LSND collaboration \cite{12}.

Before calculating the neutrino properties in more detail, we have
to discuss the contributions from the charged lepton mass
matrix and from renormalization group effects. The charged
lepton mass matrix cannot be
expected to be diagonal in the basis used. But it should resemble
the down-quark mass matrix shown in (\ref{1}). Fortunately,
because of the small mixing angles, its precise form
is not of importance at present. I just take the
suggestion for this matrix from ref. \cite{2}, transform it to our
basis and use, as an example, CP-violating phases in analogy to
$m_D$.
\be\label{10}
m_E=\left(\begin{array}{ccc}
0&-i\sqrt{\frac{3}{2}}\sigma^2&i\sigma^2\\
i\sqrt{\frac{3}{2}}\sigma^2&-\sigma&i\frac{\sigma}{\sqrt2}\\
-i\sigma^2&-i\frac{\sigma}{\sqrt2}&1\end{array}\right)
m_\tau~~.\ee
By diagonalizing  $m_E$
\be\label{11}
\quad m_E=U_E m_E^{diagonal} U_E^\dagger\ee
 the neutrino mass matrix (in the basis in which the charged 
lepton matrix is diagonal) reads
\be\label{12}
\tilde m_\nu=U^T_E m_\nu U_E ~.\ee 
Because $U_E$ is not a real matrix, CP-violation effects are
predicted. For CP-conserving processes it will turn out that 
the influence of $U_E$ on $m_\nu$ is  not essential, however. 
Before giving numerical
examples, the effects of the scale changes between the high 
scale $M_0$ and the weak scale has to be studied.

\section{Renormalization group effects}

The existence of generation quantum numbers insures the stability 
of the mass matrices against strong loop corrections. Since the 
charges of the heavy neutrinos are now fixed one can give 
corresponding charges to the up-quarks. Because of (\ref{6}) the 
singlet anti-up-quark fields may carry the 
same charges as the heavy neutrinos. 
The structure of $m_U$ then suggests that the left handed u-quark, 
charm quark and top quark fields have the 
charges 7/2~,~1/2~,~-3/2~, respectively.  

The close connection between quark and lepton mass matrices 
assumed here must have its origin at the
high scale $M_0$ which, as we have seen, is of order $10^{11}-10^{12}~
GeV$. If not before, at least at this scale new physics will set in. 
It could modify the
scale dependence of the gauge-coupling constant $g_1$ such as to
unify with $g_2$ and $g_3$ at their meeting point at $10^{16} ~GeV$. 
In any case
our task is to fix $\tilde m_\nu$ at the scale $M_0$ and to study the
behaviour of $\tilde m_\nu $ between $m_Z$ and $M_0$.
 
When applying the renormalization group
equations to the charged leptons, it is of advantage to transform
-- at all scales relevant here -- the right-handed charged leptons
such that the corresponding mass matrix contains the left-handed
mixing matrix only
\be\label{13}
\quad m_E= U_E ~m_E^{diagonal}\ee 
where $m^{diagonal}_E$ is a diagonal and positive definite real matrix.
By inserting this matrix into the
renormalization group equation, one observes that the scale
changes concern the mass eigenvalues only. $U_E$ remains
invariant: Since below $M_0$ the masses of the heavy neutrinos do 
not appear in the renormalization group equation the product 
$ U^{\dagger}_E \cdot \frac{\partial}{\partial t}U_E$ 
is a real diagonal matrix. 
This property
suffices to insure that the unitary matrix $U_E$ is
independent of the scale function
$t=\ln\mu/\mu_0$. Consequently, $U_E$ computed from
(\ref{10}), (\ref{11}) can also be used at the scale $M_0$.

At the scale $M_0$ the mass matrix $M$ for the heavy neutrinos is
obtained by replacing in (\ref{8}) $\sigma^2$ by
$\sigma^2(M_0)=m_c(M_0)/m_t(M_0)$. The mass matrix $m_\nu$ for the
light neutrinos becomes \footnote {Because of the uncertainties of the 
quark masses it is not clear whether
the relation  $m_u/m_t=(m_c/m_t)^2$ which is not strictly 
scale-invariant
but used in (\ref{2}) and (\ref{6}) holds better at $m_Z$ 
or at $M_0$.
If it holds at $M_0$, eq. (\ref{14}) and eq. (\ref{9})
(with $\sigma=\sigma(M_0)$ and
$m_t=m_t(M_0)$) are identical.}

\be\label{14}
m_\nu(M_0)=-\left(\begin{array}{ccc}
0&0&r\frac{m_u(M_0)}{m_c(M_0)}\\
0&1&-p\\
r\frac{m_u(M_0)}{m_c(M_0)}&-p&p^2\end{array}\right)
\frac{m_c(M_0)m_t(M_0)}{rM_0}~~.\ee  
\noindent
It remains to solve the renormalization group equation for
$\tilde m_\nu(\mu)$
with the boundary condition 
\be\label{15}
\tilde m_\nu(M_0)=U^T_E m_\nu(M_0)~U_E ~.\ee
According to ref. \cite{13} one has
\begin{eqnarray}\label{16}
(4\pi)^2\frac{d}{dt}\tilde m_{\nu}&=
&(-3 g^2_2+2\lambda)~ \tilde m_{\nu} +
\frac{4}{v^2} Tr(3 m_U m^{\dagger}_U+
3 m_D m^{\dagger}_D+m_E m^{\dagger}_E)~ 
\tilde m_{\nu} \nonumber \\ 
 &-&\frac{1}{v^2}(\tilde m_{\nu} m^{\dagger}_E m_E +
 m^T_E m^*_E \tilde m_{\nu} ) ~~.\end{eqnarray}
\noindent
$\lambda=\lambda(t)$ denotes the Higgs coupling constant related to 
the Higgs mass according to $m_H=\lambda v^2$ with $v=246 ~GeV$. We 
take $m_H(m_Z)=140 ~GeV$ for the numerical estimates. Eq(\ref{16}) 
simplifies since according to (\ref{12}) and (\ref{15}) $m_E$ has to be 
taken in diagonal form. Solving it gives the neutrino mass matrix 
$ \tilde m_\nu$ at the scale of the standard model. 
The neutrino mixing matrix $U=U(m_Z)$ can then be obtained by 
diagonalizing the hermitian matrix 
$\tilde m_{\nu} \cdot \tilde m^*_{\nu}$ :
\be\label{17}
\tilde m_{\nu}(m_Z) \cdot \tilde m^*_{\nu}(m_Z)=U D D^* U^{\dagger} ~.\ee
The diagonal matrix $D$ 
\be\label{18} 
D=U^{\dagger}\tilde m_{\nu}(m_Z) ~U^*  \ee
gives us the (complex) neutrino mass eigenvalues. By introducing 
the diagonal phase matrix $\Phi$ which consists of the phase factors 
of $D$ with angles divided by 2, $U$ can be redefined: 
$U \rightarrow \hat U=U \Phi$ such that (\ref{18}) gives now 
positive definite neutrino mass eigenvalues. $\hat U$ expresses 
the light neutrino 
states $ \nu_e,\nu_{\mu},\nu_{\tau}$ by the neutrino mass eigenstates 
$\nu_1,\nu_2,\nu_3$ according to
\be\label{19}
\left(\begin{array}{l} \nu_e\\ \nu_\mu\\\nu_\tau\end{array}\right)
=\hat U\left(\begin{array}{l}\nu_1\\ \nu_2\\ \nu_3\end{array}\right)~.\ee

It turns out that the mixing matrix is not strikingly different from 
the mixing matrix obtained by diagonalizing (\ref{9}) , but it 
contains CP violating phases.

\section{Results and discussions}
As shown in section 2 the mass matrices of charged fermions have a 
simple structure. We know much less about the neutrino mass matrix 
but it is tempting to assume that there exists an intimate relation 
between the up quark mass matrix and the mass matrix of the heavy 
neutrinos (the singlets with respect to the standard model gauge 
group). Because the singlet neutrino fields couple among each other, 
already the mere existence of a generation quantum number which governs 
the powers of $\sigma$ severly restricts the structure of this matrix. 
Apart from the scale $M_0$ we are left with essentially only two 
parameters ($r$ and $p$). Applying then the see-saw mechanisme we 
arrived at an interesting mass matrix for the light neutrinos 
(Eq(\ref{9})). The neutrino mass spectrum obtained from it consists 
of two nearly degenerate states which are lighter by a factor of 
order $\sigma^2$ than the third neutrino. Diagonalization of the 
neutrino mass matrix gives large mixing angles. Taking the heaviest 
mass of the light neutrino to be about $5 \cdot 10^{-2} eV$ the mass 
scale of the singlet neutrinos is of order $10^{12} ~GeV$. 
Scaling the mass matrix down from this value to the weak interaction 
scale and including also the mixings of the charged leptons, leads 
to corrections but does not change the general picture. The 
charged lepton matrix, together with the neutrino matrix, causes 
CP violating effects, however. For an illustration the 
form (\ref{10}) of the charged lepton mass matrix is used in 
the following numerical examples.

Let us start by putting the parameter $p$ equal to one. This is 
an appealing choice because of the corresponding factor one in the 
up quark mass matrix. With this value the neutrino mixing matrix 
$U$ as obtained from (\ref{17}) is of bimaximal type: Almost 
independent of the parameter $r$ the magnitudes of the elements 
of the mixing matrix are
\be\label{20}
Abs[U]= 
\left(\begin{array}{ccc}0.70& 0.71&0.05\\
0.50&0.50&0.71\\  0.50&
0.50&0.71 \end{array}\right)~.\ee

To obtain contact with the atmospheric neutrino data \cite {4} 
the product $r\cdot M_0$ can be adjusted to give the heaviest 
of the light neutrinos 
a mass of $5.5\cdot10^{-2} eV$. 
One finds $r\cdot M_0 \approx 7\cdot10^{11}~GeV$ . 
The masses of the two lighter neutrinos are 
then $ \approx r\cdot 6\cdot10^{-5}~eV$. 
The mass splitting 
between these neutrino depends on the parameter $r$ in a more involved 
way. One has e.g. $m^2_2-m^2_1$  equal 
to $0.8 \cdot 10^{-11}, ~6.5 \cdot 10^{-11},~2.2 \cdot 10^{-10}$ 
and $5.2 \cdot 10^{-10} ~({\eV})^2$ for $r=1$ , $r=2$, $r=3$ and $r=4$ , 
respectivly. 
These mass differences are in the region of 
the ones needed for the vacuum oscillation solution for solar 
neutrinos \cite {11}.

To describe the neutrino surviving and transition probabilities it is 
convenient to introduce the abbreviations
\be\label{21}
S_{ik}= \sin^2 (1.27~(m^2_i-m^2_k)\frac{L}{E})~,~~~
T_{ik}= \sin (2.54~(m^2_i-m^2_k)\frac{L}{E})  ~.\ee

The mass differences $m^2_i-m^2_k$ are taken in units of $(eV)^2$, the 
neutrino energy $E$ in MeV and $L$, the distance between generation 
and detection point in meter. The probabilities obtained from (\ref{19}) 
for $p=1$ and $r=2$ are
\begin{eqnarray}\label{22}
P(\nu_e \rightarrow \nu_e)&=&1-S_{21}-0.004 ~S_{31}-
0.004 ~S_{32}  \nonumber\\
P(\nu_{\mu} \rightarrow \nu_{\mu})&=&1-0.25 ~S_{21}-
0.51 ~S_{31}-0.49 ~S_{32}  \nonumber\\
P(\nu_{\tau} \rightarrow \nu_{\tau})&=&1-0.25~S_{21}-
0.50~S_{31}-0.50~S_{32}  \nonumber\\ 
P(\nu_e \rightarrow \nu_{\mu})&=
&0.50 ~S_{21}+0.007 ~S_{31}-0.003 ~S_{32} \nonumber\\
&+& 0.02 ~T_{21}- 0.02 ~T_{31}+0.02 ~T_{32}  \nonumber\\
P(\nu_e \rightarrow \nu_{\tau})&=
&0.50 ~S_{21}-0.003 ~S_{31}+0.007 ~S_{32}
 \nonumber\\
&-& 0.02 ~T_{21}+0.02 ~T_{31}-0.02 ~T_{32}  \nonumber\\
P(\nu_{\mu} \rightarrow \nu_{\tau})&=
&-0.25 ~S_{21}+0.50 ~S_{31}+0.49 ~S_{32}
 \nonumber\\
&+& 0.02 ~T_{21}-0.02 ~T_{31}+0.02 ~T_{32}~~.
\end{eqnarray}
Only the small numbers appearing in (\ref{22}) 
depend notably on the value of the parameter $r$.

For the solar neutrinos one can set 
$S_{31}=S_{32}=1/2, ~T_{31}=T_{32}=0$. 
For the atmospheric neutrinos, on the other hand, one can 
put $S_{21}=T_{21}=0, ~S_{31}=S_{32}, ~T_{31}=T_{32}$. 
From (\ref{22}) maximal mixing for the solar as well as 
the atmospheric neutrinos 
is obvious. It is also seen, that CP violating effects 
described by the factors multipying $T_{ik}$ are 
small in this scenario.

The bimaximal mixing obtained so far gets spoiled if the parameter $p$ 
is sizeable different from one: the mixing angle relevant for 
atmospheric neutrinos is sensitive to the value of $p$. Still, 
deviations from $p=1$ by up to 25 \% are tolerable. $p=0.75$ 
~e.g. gives for $P(\nu_{\mu} \rightarrow \nu_{\mu})$ 
and $S_{21}=0, ~S_{31}=S_{32}$

\be\label{23}
P(\nu_{\mu} \rightarrow \nu_{\mu})= 1-0.93 ~S_{31}  ~.\ee
\noindent
The Dirac neutrino matrix may differ from the up quark mass matrix.
However, if both matrices commute at the scale of $M_0$, as one 
might expect, the corresponding changes can be absorbed into 
the parameters $r$ and $M_0$. An effective parameter $r \approx 10$
would lead to a mass difference $m_2^2-m_1^2 \approx 10^{-8}~(eV)^2$ and 
thus to an energy independent suppression of solar neutrinos, in some
conflict with the results of the Homestake collaboration. 
 
\vspace{1cm}\noindent
{\bf Acknowledgement}

It is a pleasure to thank my collegues Dieter Gromes and 
Christof Wetterich for very useful comments.

\end{document}